\documentclass[aps,twocolumn,superscriptaddress,amsmath,amssymb,floatfix,showpacs,amsfonts,longbibliography]{revtex4-1}
\usepackage{times}
\usepackage[varg]{txfonts}
\usepackage{textcomp}
\usepackage{graphicx}
\usepackage{subfigure}
\usepackage{tabu}
\usepackage{color}
\usepackage[colorlinks=true,citecolor=blue,urlcolor=blue,linkcolor=blue,hyperindex,dvipdfmx]{hyperref}
\usepackage{braket}
\usepackage{float}
\usepackage{overpic}
\usepackage{amssymb}
\usepackage{amsmath}
\usepackage{gensymb}
\usepackage{mhchem}
\usepackage{mathrsfs}
\usepackage{bm}
\usepackage{diagbox}

\allowdisplaybreaks

\begin{document}

\title{Fractional and composite excitations of antiferromagnetic quantum spin trimer chains}

\author{Jun-Qing Cheng}
\thanks{These authors contributed equally to this work.}
\affiliation{State Key Laboratory of Optoelectronic Materials and Technologies, Center for Neutron Science and Technology, School of Physics, Sun Yat-Sen University, Guangzhou 510275, China}

\author{Jun Li}
\thanks{These authors contributed equally to this work.}
\affiliation{State Key Laboratory of Optoelectronic Materials and Technologies, Center for Neutron Science and Technology, School of Physics, Sun Yat-Sen University, Guangzhou 510275, China}

\author{Zijian Xiong}
\affiliation{State Key Laboratory of Optoelectronic Materials and Technologies, Center for Neutron Science and Technology, School of Physics, Sun Yat-Sen University, Guangzhou 510275, China}
\affiliation{Department of Physics, Chongqing University, Chongqing, 401331, China}
\author{Han-Qing Wu}
\email{wuhanq3@mail.sysu.edu.cn}
\affiliation{State Key Laboratory of Optoelectronic Materials and Technologies, Center for Neutron Science and Technology, School of Physics, Sun Yat-Sen University, Guangzhou 510275, China}

\author{Anders W. Sandvik}
\email{sandvik@bu.edu}
\affiliation{Department of Physics, Boston University, 590 Commonwealth Avenue, Boston, Massachusetts 02215, USA}
\affiliation{Beijing National Laboratory for Condensed Matter Physics and Institute of Physics, Chinese Academy of Sciences, Beijing 100190, China}

\author{Dao-Xin Yao}
\email{yaodaox@mail.sysu.edu.cn}
\affiliation{State Key Laboratory of Optoelectronic Materials and Technologies, Center for Neutron Science and Technology, School of Physics, Sun Yat-Sen University, Guangzhou 510275, China}

\begin{abstract}
Using  quantum Monte Carlo, exact diagonalization and perturbation theory, we study the spectrum of the $S=1/2$ antiferromagnetic
Heisenberg trimer chain by varying the ratio $g=J_2/J_1$ of the intertrimer and intratrimer coupling strengths.
The doublet ground states of  trimers form effective interacting $S=1/2$ degrees of freedom
described by a Heisenberg chain. Therefore, the conventional two-spinon continuum of width $\propto J_1$ when $g=1$
evolves into to a similar continuum of width $\propto J_2$ when $g\to 0$. The intermediate-energy and high-energy modes are termed \emph{doublons}
and \emph{quartons} which fractionalize
with increasing $g$ to form the conventional spinon continuum. In particular,
at $g \approx 0.716$, the gap between the low-energy spinon branch and the
high-energy band with mixed doublons, quartons, and spinons closes.
 These features should be observable in inelastic neutron scattering experiments
if a quasi-one-dimensional quantum magnet with the linear trimer structure and $J_2<J_1$ can be identified.
Our results may open a  window for exploring the  high-energy fractional excitations.
\end{abstract}

\maketitle
\date{\today}


\section{\label{sec:level1} Introduction}

Many quasi one-dimensional (1D) magnetic materials with spin $S=1/2$ moments harbor exotic phenomena originating from the physics of the Heisenberg
antiferromagnetic chain (HAC) and its extensions \cite{Mikeska2004}. The dynamic spin structure factor of \ce{KCuF_3} \cite{tennant93,lake13},
measured using inelastic neutron scattering, exhibits the characteristic gapless two-spinon continuum \cite{karbach97} of the uniform HAC. A phase
transition to a gapped dimerized state, driven by additional frustration and spin-phonon couplings has been realized in \ce{CuGeO_3}
\cite{hase93}. In systems with random couplings, the random-singlet state with infinite dynamic exponent forms \cite{fisher94,shu16}, as originally
observed in a class of Bechgaard salts \cite{bulaevskii72,azevado77} and more recently in \ce{BaCu_2SiGeO_7} \cite{shiroka2013} and
\ce{BaCu_2(Si_{0.5}Ge_{0.5})_2O_7} \cite{masuda2004}. Furthermore, the resonant inelastic x-ray scattering (RIXS) technique has now enabled specific
detection of multi-spinon excitations \cite{PhysRevLett.106.157205,RIXS2018NC} in the HAC material \ce{Sr_2CuO_3}, and string excitations
have been identified by terahertz spectroscopy in the Heisenberg-Ising compounds \ce{SrCo_2V_2O_8} \cite{Wang2018219} and \ce{BaCo_2V_2O_8}
\cite{PhysRevLett.123.067202}.

A unit cell of more than one spin, which is the context of our work presented in this paper, can lead to an even richer variety of  1D magnetic properties\cite{dagotto96,doretto09,PhysRevB.103.184415}. For example,
ladder systems  with rungs consisting of an odd or even number of $S=1/2$ spins
have a gapless or gapped spectrum, respectively \cite{dagotto96}, in a way similar
to Haldane's conjecture \cite{haldane1983} of chains with  half-odd-integer or integer spins.
Trimer chains have also been studied experimentally, including \ce{A_3Cu_3(PO_4)_4}
(\ce{A=Ca, Sr, Pb}) \cite{PhysRevB.71.144411,drillon19931d,belik2005long,PhysRevB.76.014409} and \ce{(C_5H_11NO_2)_2.3CuCl_2.2H_2O}
\cite{hasegawa2012magnetic}, where in all cases the structure is such that two spins in one trimer are coupled to two spins of their
neighboring trimers. The linear trimer chain with repeated couplings $J_1$-$J_1$-$J_2$ (intratrimer $J_1$ and intertrimer $J_2$),
is realized in \ce{Cu(P_2O_6OH)_2} \cite{PhysRevB.73.104419} and \ce{Cu(P_2O_6OD)_2} \cite{PhysRevB.76.064431}, both of which have
$J_2>J_1$. The case of strong intertrimer coupling has also been investigated theoretically \cite{PhysRevB.103.184415}, and other quantum
magnets with trimerized structure have also attracted attention\cite{Cao2020,PhysRevResearch.2.033382,PhysRevB.103.125120}.

\begin{figure}[t]
\includegraphics[width=8cm]{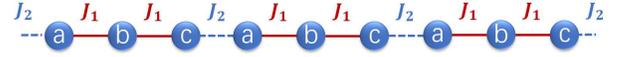}
\caption{\textbf{Schematic representation of a trimer spin chain}. $J_1$ and $J_2$ represent the two different antiferromagnetic nearest-neighbor Heisenberg intratrimer  and intertrimer couplings, respectively.  We here consider systems with $J_1 \geq J_2 >0$. We will use the letters $a,b,c$ as indicated to
refer to the three spins within a given unit cell.}
\label{figure1}
\end{figure}

From the theoretical perspective, it is interesting to consider the linear trimer chain illustrated in Fig.~\ref{figure1}, where for
$J_2 \ll J_1$ the excitations can be understood from perturbative calculations starting from the eigenstates of the isolated trimers.
Surprisingly, though the case $J_2>J_1$ has been studied both experimentally and theoretically, the potential of the $J_1>J_2$ system
($J_1,J_2$ both antiferromagnetic) to realize a host of interesting excitations and their confinement---deconfinement cross-overs has not
been recognized in the previous literature. We will study this model system extensively here.

An interesting preliminary aspect of the $J_1$-$J_1$-$J_2$ system, where we define $g\equiv J_2/J_1 \in (0,1]$, is
that it reduces to the conventional HAC when $g=1$, while for $g \ll 1$ the low-energy excitations can be mapped onto an HAC consisting
of one effective $S=1/2$ spin in each unit cell. Thus, in both these limits the low-energy excitations should be spinons, but they live in different
Brillouin Zones (BZs). By reducing the coupling ratio from $g=1$ one can expect an evolution from the conventional two-spinon continuum in the window
$q \in [0,\rm{\pi}]$ of width $\propto J_1$  into
three continua in the windows $q \in [0,\rm{\pi}/3]$, $[\rm{\pi}/3,2\rm{\pi}/3]$, $[2\rm{\pi}/3,\rm{\pi}]$ with band width $\propto J_2$. Moreover, at small values of $g$ there should
also be weakly dispersive modes arising from the internal excitations of the trimers, and these must eventually fractionalize into spinons as $g \to 1$.
The interplay between the two types of spinons and the higher-energy modes, and how these coexisting excitations eventually evolve into just the conventional
spinon continuum, is not immediately clear.
Our calculations reported here are also in part motivated by recent work on two-dimensional systems consisting of weakly coupled
multi-spin plaquettes of different shapes \cite{PhysRevB.99.085112,PhysRevB.99.174434}. In the calculations for coupled $3\times 3$ plaquettes,
intriguing spectral features were found but were not fully explained \cite{PhysRevB.99.085112}, and the simpler 1D system considered here can guide
additional calculations and provide interpretations.

 We will develop a physical understanding of the various observed branches of excitations and their intricate evolution with $g$
by interpreting the numerical spectral functions  in the light of perturbative calculations  as well as the known properties of
the spinon continuum in the uniform HAC.
The $S=1$ excited states  of the uniform $S=1/2$ HAC are fractionalized into independently propagating particles, spinons, each carrying $S=1/2$ \cite{cloizeaux62,Bethe,faddeev81}. The leading contributions of these excitations (two-spinon contributions) with
total momentum $q$ to the dynamic structure factor
$\mathcal{S}(q,\omega)$ can be calculated relatively easily \cite{karbach97} by Bethe ansatz (BA) calculations, while four-spinon contributions
require sophisticated numerical calculations with the BA states \cite{caux06}. The utility of the BA is limited, however, when perturbing
the HAC beyond the solvable  XXZ model \cite{caux06,wu19}. In general reliable calculations of dynamical properties
are very challenging beyond the small lattices accessible to exact diagonalization (ED) techniques \cite{dagotto1996}.

Currently,  the  density matrix renormalization
group (DMRG) \cite{PhysRevLett.69.2863,PhysRevLett.93.076401} and related methods formulated with matrix-product states (MPS) \cite{SCHOLLWOCK201196} are very powerful for 1D systems and are also applicable for calculations of dynamical structure factors\cite{PhysRevB.77.134437,PAECKEL2019167998},
primarily for systems with open boundaries, due to inefficiency in the case of periodic chains.
Quantum Monte Carlo (QMC) simulations with subsequent numerical analytic continuation of the imaginary-time correlation functions \cite{PhysRevB.57.10287,beach2004identifying,PhysRevB.78.174429,SSE1,SAC1,SAC2,SAC3} can be applied to large periodic system sizes in any number of dimensions as long as the negative sign problem
can be avoided. Some very useful results
for $\mathcal{S}(q,\omega)$ have been obtained for a variety of quantum magnets, e.g., in
Refs.~\cite{PhysRevLett.122.127201,SAC2,SAC3,PhysRevLett.118.147207,PhysRevB.98.174421,PhysRevLett.120.167202}.

We here study the trimer chain using both the QMC  and ED methods. For the former, we compute imaginary-time correlations with the stochastic
series expansion \cite{SSE1} QMC method and employ a variant of the stochastic analytic continuation (SAC) technique
\cite{PhysRevB.57.10287,beach2004identifying,PhysRevB.78.174429,SAC1,SAC2,SAC3}. We will discuss results for $\mathcal{S}(q,\omega)$
in both the regular and reduced BZs. When $g$ is small, three well separated features
are observed. A low-energy continuum extending up to $\omega \propto J_2$ with a characteristic spinon structure is present
due to the fact that each trimer hosts an effective spin-$1/2$ degree of freedom and an effective HAC with coupling $\propto J_2$ forms. At higher energies,
$\omega \propto J_1$, there are two different weakly dispersing modes corresponding to the internal excitations of the trimers.  These  modes evolve
significantly as $g$ is increased, and eventually for $g \to 1$ the standard spinon continuum of the uniform chain is recovered.
 In order to
further understand the excitation mechanism and the properties of these excitations, we construct perturbatively motivated expressions for propagating
internal trimer excitations and find excellent agreement with the numerical results for small to moderate values of $g$.  According to the characters of
these excitations, we call the quasiparticle corresponding to the intermediate-energy excitation ($\omega \approx J_1$) the {\it doublon} and the one corresponding
to the higher energies ($\omega \approx 1.5J_1$) the {\it quarton}. These excitations may be helpful for understanding similar dual flat bands observed in the inelastic neutron-scattering spectra of \ce{A_3Cu_3(PO_4)_4} (\ce{A=Ca, Sr, Pb}) \cite{PhysRevB.71.144411}, where there are  next-nearest-neighbor interactions but where trimers are also effectively weakly coupled as in the $g \ll 1$ system studied here.

\begin{figure*}
	\includegraphics[width=16cm]{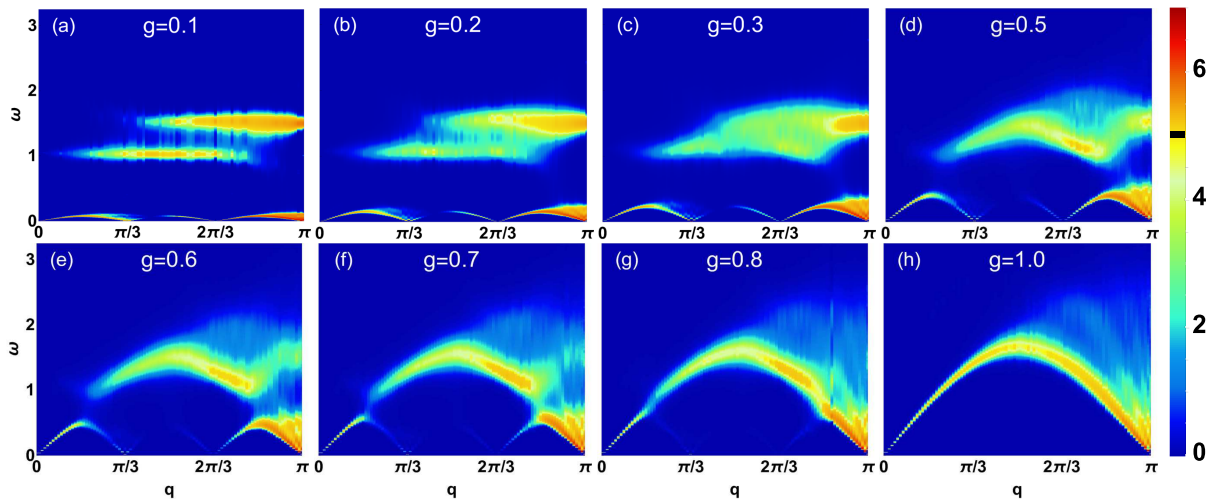}
	\caption{\label{figure2} \textbf{Dynamic spin structure factor $\mathcal{S}(q,\omega)$ obtained from QMC-SAC calculations for different
		$g$ values}.  The  color coding of $\mathcal{S}(q,\omega)$, illustrated with the bar on the right side, uses a piecewise function
		where the boundary value $U_0=5$ is indicated by the line on the color bar.
		Below the boundary, the low-intensity portion is characterized by a linear mapping of the spectral
		function to the color bar, while above the boundary a logarithmic scale is used, $U=U_0+\log_{10}[\mathcal{S}(q,\omega)]- \log_{10}(U_0)$. Since the SAC method generates spectra with intrinsic broadening, no additional
		broadening is imposed here.
		The vertical striped features,
		noticeable especially between the $\omega \approx 1$ and $\omega \approx 1.5$ bands in (a) and (b), are typical in analytic continuation when the statistical
		precision of the QMC imaginary-time data is barely sufficient for resolving two features, thus not completely resolving them for some momenta while resolving
		them in other cases (c)-(h).}
\end{figure*}

We will also use ED to compute the dynamic structure factor in a truncated  Hilbert space, where the full low-energy
spinon space is included in addition to one each of the doublon and quarton. For $g \alt 0.5$, the good agreement of the spectral functions obtained in this manner with
those from the other calculations confirm the nature of these excitations, while disagreements for larger $g$ show how the doublons and quartons loose their identity
when they begin to fractionalize into the standard HAC spinon continuum that emerges when $g \to 1$. In the intermediate $g$ regime, we have two different spinon continua
coexisting with the doublon and quarton.
 These calculations demonstrate how two
branches of trimer excitations gradually broaden out when $g$ increases, then merge together and evolve into the upper part of the two-spinon continuum in a
fractionalization process.  Based on these results, we also develop an intuitive picture of the quasi-particles as complex domain walls, generalizing
the standard domain-wall description of spinons in the HAC. The high-energy spin excitations (energy of order $J$) have also been under intense
scrutiny in the antiferromagnetic parent compounds of the high-$T_c$ superconductors
\cite{PhysRevLett.105.247001,Zhou2013,Ishii2014,PhysRevB.91.184513,Song2021} and other systems described by the 2D Heisenberg model
\cite{piazza15,SAC2}. Our results on the fractionalization mechanism of high-energy spin excitations of the trimer chain may be helpful for
further exploring the fractionalization mechanism of high-energy magnons in these systems.

\section{\label{section2} Results}

\subsection{\label{section2-A}Model}

The Hamiltonian of the spin-$1/2$ antiferromagnetic trimer chain with periodic boundary conditions reads
\begin{equation}
H=\sum_{i=1}^{N}\left[ J_1 \left(\mathbf{S}_{i,a}\cdot \mathbf{S}_{i,b} +\mathbf{S}_{i,b} \cdot \mathbf{S}_{i,c} \right)
+ J_2 \mathbf{S}_{i,c} \cdot\mathbf{S}_{i+1,a} \right],
\end{equation}
where $\mathbf{ S}_{i,\alpha}$ is the spin-$1/2$ operator at the $\alpha$th site in the $i$th trimer, the intratrimer labels $\alpha \in \{a,b,c\}$ are explained
by Fig.~\ref{figure1}. The total number of trimers is $N$ and the total
length of system is $L=3N$. The tuning parameter $g$ is defined as $g = J_2/J_1$ and we here limit our study to $0 \le g\le 1$. For simplicity, we set the
intratrimer interaction $J_1=1$ as the energy unit, so that intertrimer interaction $J_2=g$. Our interest is in the whole range of intermediate
coupling ratios $g \in (0,1)$ where the system evolves between the isolated trimers and the isotropic BA solvable HAC.

The dynamic structure factor describing the time dependent spin-spin correlations at a given transferred momentum $q$ is defined as
\begin{eqnarray}
	\label{Sqw}
	\mathcal{S}^{ \beta \gamma} (q,\omega) =
	\frac{1}{2\rm{\pi}} \int_{-\infty}^{\infty}dt \left\langle S^\beta_q (t) S^\gamma_{-q}(0) \right\rangle  \rm{e}^{\rm{i}\omega \emph{t}},
\end{eqnarray}
where  $\beta$,$\gamma$ refer to spin components $x,y,z$, and $S_{q}^{\beta(\gamma)}$ is the Fourier transform of the spins that we discuss later
(as it depends on the type of BZ used). In frequency space the dynamic structure factor is
\begin{equation}
	\mathcal{S}^{zz} (q,\omega)=
	\sum_{n}\left|\left\langle \psi_n |S_{q}^z|\psi_0\right\rangle \right|^2 \delta\left[\omega-(E_n-E_0)\right],
	\label{sqwdef}
\end{equation}
where we have explicitly indicated the diagonal $z$ component, which already contains all information in the case of the spin-rotational invariant
model considered here. From now on we will omit the superscripts and define $\mathcal{S} (q,\omega) \equiv 3\mathcal{S}^{zz} (q,\omega)$.


\subsection{\label{section2-B}Overview of numerical results}

\begin{figure*}[t]
	\includegraphics[width=14cm]{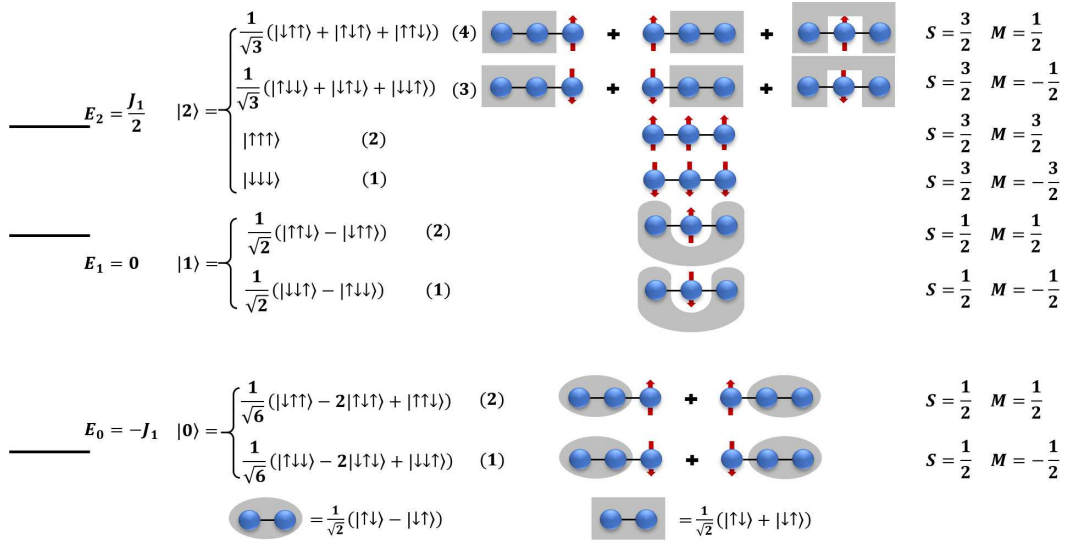}
	\caption{\label{figure3}  \textbf{The level spectrum  and corresponding wave functions of one isolated trimer}. The second column lists the wave
		functions in the spin-$z$ basis, while the third column presents the spin structures using a basis of singlets (grey ovals and rounded shapes),
		zero-magnetization triplets (grey square shapes), and unpaired spins (arrows). The rightmost column lists the total spin quantum number $S$ and
		magnetic quantum number $M$. }
\end{figure*}

In this section, we present the spin excitation spectra of the trimer chain obtained by  QMC-SAC calculations (see the technical details in Methods \ref{methods}).
Color plots of  $\mathcal{S}(q,\omega)$ representing the full $(q,\omega)$ space are shown in  Fig.~\ref{figure2}  for a chain with $192$ spins.
We have also calculated the full spectrum by ED for a chain with $30$ spins (see  Supplementary Figure~1), which exhibits similar features as the
QMC-SAC results but with significant finite-size effects. To better reveal the important spectral features, here and in graphs in later sections,
we used color coding of $\mathcal{S}(q,\omega)$ by a piecewise function which is explained further in the caption of Fig.~\ref{figure2}.

Let us first examine results for $g=1$, shown in Fig.~\ref{figure2}(h), where we observe the well understood characteristic asymmetric spinon
continuum of the HAC chain. As $q\to 0$, the width of the continuum vanishes along with the spectral weight, while at the other gapless point $q=\rm{\pi}$ the continuum
has maximum width and the weight in the thermodynamic limit is $\omega^{-1}$ divergent (with a logarithmic correction). At the lower edge away from the
gapless points, the dispersion relation $\epsilon_q = (\rm{\pi}/2)|\sin(q)|$ reflects that of a single spinon. The spectral weight is also concentrated
at this edge, with a $(\omega-\epsilon_q)^{1/2}$ divergence. The divergent features can of course not be strictly observed in the finite systems, and in the case
of QMC-SAC results there is also broadening and some distortions due to the incomplete data used in the analytic continuation. Nevertheless,
the lower spectral bound is well reproduced and the observed concentration of spectral weight at the lower edge is a true feature
of the spinon continuum \cite{karbach97,caux06}.

Next, we discuss how the spectral function evolves as $g$ is increased from $0$ and eventually approaches $1$. As apparent in Fig.~\ref{figure2}(a)-(c), when $g=0.1-0.3$ the gapless low-energy spectrum comprises spinon excitations originating from an effective HAC of $N$ effective
$S=1/2$ degrees of freedom.
The reduced BZ corresponds to $q \in [0,\rm{\pi}/3]$ in the full BZ used in the figures. The same excitations appear also in the
windows $q \in [\rm{\pi}/3,2\rm{\pi}/3]$ and $q \in [2\rm{\pi}/3,\rm{\pi}]$, with different weight distributions due to the different phase factors. In addition to $q=0,\rm{\pi}$, the spectrum is gapless at $q=\rm{\pi}/3$ and $q=2\rm{\pi}/3$ when $g<1$, which can also be explained by the Lieb-Schultz-Mattis  theorem \cite{LIEB1961407,tasaki2020physics} along with BZ folding effect since the system is rotationally invariant and translationally invariant  with unit cell  containing three spins.

Many of the observed features follow from the fact that the three low-energy spinon continua must evolve into a single continuum as $g \to 1$. Thus, the spectral weight in the central portions $q \in [\rm{\pi}/3,2\rm{\pi}/3]$  of the low-energy spinon continuum decreases while the leftmost $q \in [0,\rm{\pi}/3]$ and rightmost $q \in [2\rm{\pi}/3,\rm{\pi}]$ half
arches become more prominent. A very interesting aspect of the evolution is how the intermediate-energy and high-energy modes gradually morph into the high-energy part of the standard HAC spinon
continuum. Thus, a fractionalization of the quasiparticles takes place. In the following, we will provide a perturbative analysis to explain the
evolution of the spectrum.

\subsection{\label{section2-C}Perturbative analysis: Effective Heisenberg coupling}

\begin{figure*}
\includegraphics[width=16cm]{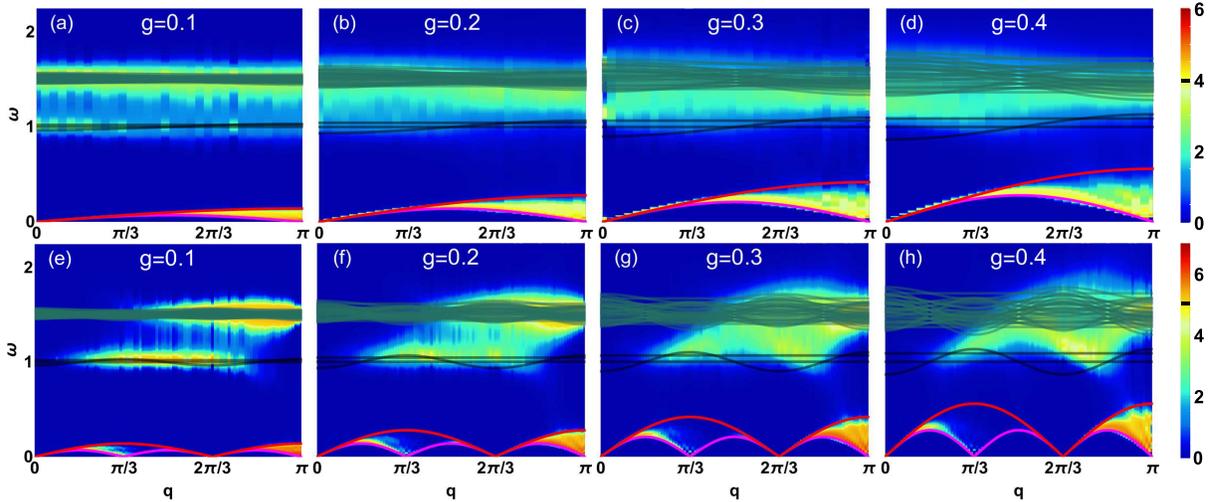}
\caption{\label{figure4} \textbf{QMC-SAC results compared with perturbative energy levels}. (a)-(d) show results
for $S_{\rm{red}}(q,\omega)$ in the reduced BZ, while (e)-(h) show $S(q,\omega)$ in the full BZ as in
Fig.~\ref{figure2}.  In (a)-(d) the  magenta and red solid lines  are respectively
the lower boundary $\omega_{\rm{l}} = \rm{\pi} J_{\rm eff} \left|\sin (q) \right|/2$ and upper boundary $\omega_{\rm{u}} =\rm{\pi} J_{\rm{eff}} \left|\sin (q/2)\right|$ of the two-spinon continuum, the black solid lines represent the
dispersion relations [see Eqs.~(\ref{intermediate2}) - (\ref{intermediate4})] corresponding to the intermediate-energy spectrum. The green solid lines are
the  dispersion relations [see Eqs.~(\ref{highfirst}) - (\ref{highlast})] corresponding to the higher-energy spectrum. The  dispersion relations in full BZ are obtained by
unfolding the results in the reduced BZ. The boundaries between the linear and logarithmic color mappings are (a)-(d) $U_0=4$ and (e)-(h) $U_0=5$, respectively, as indicated on the color bars.}
\end{figure*}

In Fig.~\ref{figure3} we  observe that the low-energy doublets  contain a singlet and an unpaired spin; thus, each trimer contains an effective
spin-$1/2$ degree of freedom. Furthermore, there is a clear separation to the higher-energy states, which according to our results in the previous
section survives at least for $g\le 0.4$. Therefore, the low-energy excitations of trimer chain can be well described by an effective Hamiltonian
whose excitations only contain the spinons. The trimer chain is translationally invariant with a unit cell of three spins and  rotationally
invariant, and the effective Hamiltonian must conserve the translation symmetry and SU(2) symmetry. We explicitly derive the effective HAC
arising from the doublet trimer ground states by applying the Kadanoff method \cite{drell1977quantum,jullien1978zero,RealSpace1996,PhysRevA.77.032346},
projecting the Hamiltonian onto the a low-energy subspace constructed from  the lowest eigenstates of each trimer (see Supplementary Note~2). The
effective Hamiltonian for $N$ trimers is
\begin{equation}
	\label{effective H}
	H_{\rm{eff}}=J_{\rm{eff}} \sum_{j=1}^{N} \widetilde{\mathbf{S}}_{j} \cdot \widetilde{\mathbf{S}}_{j+1},
\end{equation}
which describes an isotropic HAC with an effective coupling strength  $J_{\rm eff} = 4J_2/9$, and $\widetilde{\mathbf{S}}$ is the effective
spin-$1/2$ operator.
 We next define a dynamic structure factor in the reduced BZ as follows
\begin{equation}
  \label{sqomegared}
 \mathcal{S}_{\rm{red}} (q,\omega)=\sum_{\alpha=a,b,c} \mathcal{S}^{zz}_{\alpha \alpha} (q,\omega),
\end{equation}
where $\mathcal{S}^{zz}_{\alpha \alpha} (q,\omega)$ is the dynamical structure factor as defined in Eq.~(\ref{Sqw}) but including only the spins
at location $\alpha \in \{a,b,c\}$. The momenta are then of the form $q = 6n\rm{\pi}/L$ for the system of $L=3N$
spins.  The low-energy part is the two-spinon continuum, with the predicted lower boundary
$\omega_{\rm{l}} = \rm{\pi} J_{\rm{eff}} \left|\sin (q) \right|/2$ and upper boundary $\omega_{\rm{u}} =\rm{\pi} J_{\rm{eff}} \left|\sin (q/2)\right|$, is indicated in the reduced BZ in Figs.~\ref{figure4}(a)-(d). The predicted lower boundary
$\omega'_{\rm{l}} = \rm{\pi} J_{\rm{eff}} \left|\sin (3q) \right|/2$ and upper boundary $\omega'_{\rm{u}} =\rm{\pi} J_{\rm{eff}} \left|\sin (3q/2)\right|$ in the full BZ is similarly shown in Figs.~\ref{figure4}(e)-(f).
We observe that the boundaries are in good agreement with the numerical results, and  the spectral weight
close to the upper bound is very small at $q=\rm{\pi}$, as is well known in the case of the standard HAC.
Thus, we have confirmed that the trimer chain reduces to an effective HAC with exchange
interaction $J_{\rm eff}=4J_2/9 $ even for $g$ as high as about $0.4$.

\subsection{\label{section2-D}Perturbative analysis: Propagating internal trimer excitations}

Looking at the full level spectrum and corresponding eigenvectors of one single trimer in  Fig.~\ref{figure3}, the ground state is a
doublet with energy $E_0 = -J_1$, total spin quantum number $S=1/2$, and magnetic quantum number $M=\pm 1/2$.  The first and second excited states of the trimer are a doublet with $E_1=0$, $S=1/2$ and a quartet
with $E_2 = J_1/2$, $S=3/2$, respectively.  We also show the
structure of the eigenstates when written as pairs of spins forming a singlet or a triplet, with one or three left over unpaired spin.  The two low-energy doublets both contain only singlets and unpaired spins, while the
higher quadruplet excitations contain either a triplet pair and an unpaired spin or three unpaired spins. The magnetic quantum number $M=\pm 1/2$ or $M=\pm 3/2$
corresponds to the number of unpaired spins in each case.

Taking two trimers as an example, when $g$ is small enough the coupling can be regarded as a perturbation of the product state of the
isolated trimers. There are four  states forming a
singlet ground state and  a triplet excitation with a gap of order $g$. In addition, the lowest internal excitations of the trimers have a gap of order
$J_1$ to the low-energy states. The almost degenerate excitations originating from the two different trimers include $S=1$ states, which are of our primary
interest when considering the dynamic spin structure factor.
Increasing the number of trimers, the analogy of the lowest singlet and triplet of the two-trimer system is the spinon continuum, which for
non-interacting spinons come with $S=0$ and $S=1$. The internal excitations of the trimers will form weakly dispersive bands for small $g$, and lose their single-trimer identity for larger $g$. For simplicity, the calculations involving these higher excitations will
not be carried out with the spin rotation symmetry maintained. Being interested in $S=1$ excitations probed with the dynamic structure factor, we will still
focus on excited states with $|\Delta M|=1$.

\begin{figure}[t]
	\includegraphics[width=8cm]{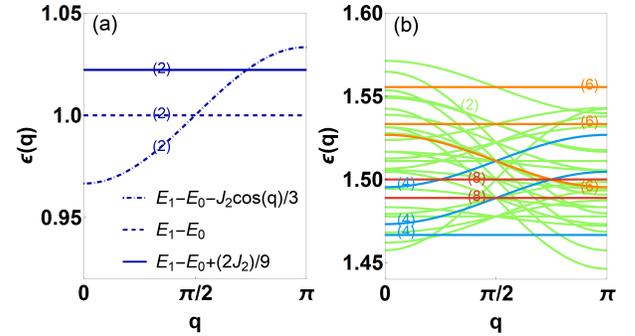}
	\caption{\label{figure5}  \textbf{Dispersion relations  of the intermediate-energy and high-energy excitations in the reduced BZ}. All results are from the case where $g=0.1$. The numbers marked on curves represent the degeneracies. In (a) each dispersion relation  is two-fold degenerate. In (b) four colors have been used to distinguish  the different degeneracies.}
\end{figure}

The internal excitations of the trimer chain are almost localized when $g$ is small, and cannot be classified as  magnons or triplons.
In order to understand the nature of these excitations, we propose a scheme to calculate their dispersion relations. The results are shown in
Fig.~\ref{figure4} as a number of dispersion relations corresponding to propagation of internal trimer excitations in the background of a chain with defects
mimicking the complex ground state of the effective HAC. The overall good agreement with the QMC--SAC results on the location of these excitations and their
band widths suggest that our picture of the excitation is correct even though the calculation involves a very rough approximation of the ground state.
The bending of, in particular, the intermediate band, and the broadening for $g\alt 0.3$ visible in the full BZ are not captured by the simple ansatz used here, which will be
 presented in detail below. The approach nevertheless forms a good starting point before considering other approaches.

By assuming  the ground-state wave function of the spin chain is the product states of ground states of each trimer and the excited-state wave function contains one excited trimer,  we are able to calculate the dispersion relations corresponding to the intermediate-energy excitations  with $|\Delta M|=1$. Considering the massively degenerate states obtained from the possible choices of the doublet ground state on each trimer and
total magnetization $M=0$, it is found that the dispersion relations are mainly dependent on the excited trimers and their neighbours. Translated
to finite size, this means that the correct generic result is already obtained from a system with $N=4$ trimers. The details can be found in Supplementary Note~3. The result is six remaining dispersion relations,
\begin{mathletters}
	\begin{eqnarray}
		&&E_{1}-E_{0}-\frac{1}{3}J_{2} \cos{q} \label{intermediate2}, \\
		&&E_{1}-E_{0} \label{intermediate3},\\
		&&E_{1}-E_{0}+\frac{2}{9}J_{2} \label{intermediate4},
	\end{eqnarray}
\end{mathletters}
where each case is two-fold degenerate.
All three dispersion relations  for $g=0.1$ are graphed in Fig.~\ref{figure5}(a). Two  dispersion relations are independent of
$q$ since the excitations are localized. According to the  characters of this excitation starting from the $g=0$ limit, we refer to it
as the doublon.

Calculating the dispersion of the high-energy excitations evolving from the trimer quartet at $\omega = 3J_1/2$ is more complicated but we follow the
same strategy as for the doublon. The intermediate-energy band is formed by reconfiguring the singlet bond in the original ground state of
the trimer in Fig.~\ref{figure3} at a cost of $J_1$. The higher  band is formed out of trimers excited into their $S=3/2$ highest state, which
can be seen as the singlet of two spins excited into a triplet, at a cost of $3J_1/2$. Accordingly,  we refer to this high-energy
quasiparticles as the quartons. This rough estimation of the quarton energy  matches with the QMC-SAC results for small $g$ values.

Our calculation does not conserve the total spin of the collective many-body state, but we consider the excitations corresponding to $\Delta S=1$ on
one trimer.  As a result, $96$ dispersion relations fall into $33$ different
cases,
\begin{eqnarray}
	&&E_2-E_0-\frac{2 \sqrt{2}+1}{9} J_2 \cos (q) + \left\{0,-\frac{J_2}{9}\right\} \label{highfirst},\\
	&&E_2-E_0+\frac{2 \sqrt{2}+1}{9} J_2 \cos (q) + \left\{0,\frac{J_2}{9}, \frac{2 J_2}{9}\right\},\\
	&&E_2-E_0-\frac{\sqrt{2}}{9}  J_2 \cos (q) + \left\{\pm \frac{ J_2}{9},\pm \frac{2 J_2}{9}\right\},\\
	&&E_2-E_0+\frac{\sqrt{2}}{9}  J_2 \cos (q) + \left\{\pm \frac{ J_2}{9}, \frac{J_2}{3}, \frac{5 J_2}{9}\right\},\\
	&&E_2-E_0+\left\{0, \pm\frac{J_2}{9}, -\frac{2 J_2}{9},\pm\frac{J_2}{3},\frac{5 J_2}{9} \right\}, \label{independent}\\
	&&E_2-E_0-\frac{\sqrt{3}}{18}  J_2 \cos (q)+ \left\{-\frac{ J_2}{9},\pm \frac{2 J_2}{9},\frac{ J_2}{3}\right\},\\
	&&E_2-E_0+\frac{J_2 }{18} \cos (q)+ \left\{0,\pm\frac{ J_2}{9},-\frac{2 J_2}{9}\right\},\\
	&&E_2-E_0+\frac{J_2 }{6} \cos (q)+ \left\{0,\frac{ J_2}{9},\frac{2 J_2}{9},\frac{J_2}{3}\right\},\\	
	&&E_2-E_0+\frac{2 \sqrt{2}-1}{9} J_2 \cos (q)+\frac{J_2}{9},\label{highlast}
\end{eqnarray}
which are  displayed in Fig.~\ref{figure5}(b) for $g=0.1$. The calculation details can be found in  Supplementary Note~3. The number marked on each curve shows the degeneracy of every dispersion
relation of the high-energy excitation depending on the number of times it appears among the total $96$ cases. Among these dispersion relations,
  some are independent of $q$, see Eq.~(\ref{independent}), and the dispersion relations $
E_2-E_0-J_2/9$ and $E_2-E_0$ both have the maximum  degeneracy, eight.  It can be found that most of the dispersion relations in Eq.~(\ref{independent})   also have large degeneracies.
 The reason is that when $g$ is small,  these excitations are localized in the trimers and dominate  the whole types of excitation in our perturbative calculation. When $g$ is increased,
  the  dispersion relations dependent of $q$ will become more significant, therefore the deviation of our perturbative calculation on the condition of small $g$ will be more obvious.

\begin{figure}
	\includegraphics[width=8.5cm]{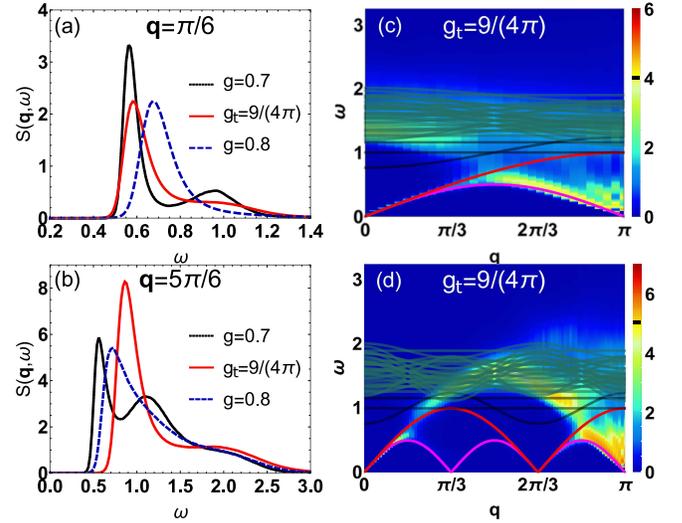}
	\caption{\label{figure6} \textbf{QMC-SAC results and perturbative energy levels  near the threshold value of $g_{\rm{t}}$}. $\mathcal{S}(q,\omega)$ from QMC-SAC calculation
at the special $q$ points (a) $q=\rm{\pi}/6$
		and (b) $q=5\rm{\pi}/6$ for $g=0.7,9/(4\rm{\pi}),0.8$, in the region where all the different excitations merge together and their individual identities as spinons, doublons, and quartons are lost.   (c) and (d) show $\mathcal{S}_{\rm{red}}(q,\omega)$  and $\mathcal{S}(q,\omega)$
		for $g_{\rm{t}}=9/(4\rm{\pi})$ in the reduced and full BZs, respectively.  The upper boundary of the
		two-spinon continuum (red solid line) and the dispersion relation Eq.~(\ref{intermediate3}) corresponding to the intermediate-energy branch (black solid
		line) cross over at $q=\rm{\pi}$.  The boundaries between the linear and logarithmic color mappings are (c) $U_0=4$ and (d) $U_0=5$, respectively, as indicated on the color bars.}
\end{figure}

\subsection{\label{section2-F}Comparisons of numerical and analytical results}

Next, we compare the above doublon and quarton dispersion relations with QMC-SAC results in both the reduced and full BZs.
At $g=0.1$, Fig.~\ref{figure4}(a), we clearly observe three different bands of excitations; in addition to the low-energy spinon continuum we have weakly
dispersive intermediate-energy ($\omega \approx J_1=1$) and higher-energy bands ($\omega \approx 2J_1 /3 = 1.5$) exactly in the regions where the QMC-SAC results
exhibit large spectral weight. Fig.~\ref{figure4}(e) shows the unfolded dispersion relations and QMC-SAC results in the full BZ. The intermediate-energy and high-energy modes are more clearly visible in the full BZ, as the three $q$ windows
have different weighting for the $a$, $b$, and $c$ trimer spin operators, thus offering more opportunities for an optimal weighting that makes any of the features
visible.  The advantages of the full BZ are even more clear at $g=0.2$ in Figs.~\ref{figure4}(b) and \ref{figure4}(f), where the dynamic structure factor
in the reduced BZ does not exhibit two separate modes but they have merged into a single continuum. This merger of the two modes may partially be due to the
limitations of the QMC-SAC approach, but most likely it reflects to a large extent the actual weight distribution.
As $g$ is increased, these two bands merge into each other also in the full BZ. It is remarkable how flat the bands are even at $g$ as large as
$0.4$, and that the perturbative calculation at least gives the correct region of dominant spectral weight in the reduced BZ. However, from the
full BZ results it is also clear that non-perturbative effects set in, with dispersive modes visible in between the two bands predicted by the
variational states with only a single excited doublon or quarton. These dispersive modes that grow out of the doublons and quartons
eventually evolve into the upper part of the spinon continuum as $g \to 1$, as seen in Fig.~(\ref{figure2}).

We can also see in Fig.~\ref{figure2} that the low-energy and high-energy bands merge near $g=0.7$.  As shown in Figs.~\ref{figure6}(a)(b),
$\mathcal{S}(\pi/6,\omega)$  and $\mathcal{S}(5\pi/6,\omega)$  both exhibit two peaks when $g=0.7$, while in both cases only one peak is present
for $g=0.8$. Along with Figs.~\ref{figure2}(f) and \ref{figure2}(g), we can conclude a threshold value $g_{\rm{t}}$  between $g=0.7$ and $g=0.8$ where the
spinon upper bound touches the lower edge of the higher-energy band. This point signifies a hybrid of the doublon and quarton excitations becoming
part of the conventional spinon continuum, which should be associated with a fractionalization mechanisms. Beyond the threshold value,
the spectrum exhibits a continuum with a single peak, tending to the standard two-spinon continuum of the
isotropic HAC when $g \to 1$.

We can derive the threshold value for the merger of the different quasiparticle bands using the perturbative dispersion relations,
which amounts to solving the equation $\pi J_{\rm eff}  \left|\sin (q/2)\right|= E_1 - E_0 $, with the result $g_{\rm{t}}=9/(4\pi)\approx0.716$.
While we do not expect this value to be very precise,
in Figs.~\ref{figure6}(a)(b),  $\mathcal{S}(q,\omega)$ at $q =\pi/6$ and $q =5\pi/6$ with $g_{\rm{t}}=9/(4\pi)$ are seen
to contain a single broad peak. Figs.~\ref{figure6}(c) and (d)
present $\mathcal{S}_{\rm{red}}(q,\omega)$ and $\mathcal{S}(q,\omega)$ obtained from QMC-SAC calculations for $g_{\rm{t}}=9/(4\pi)$,
where the  spectrum begins to form a single band. The gap between the upper boundary of the two-spinon continuum  and the dispersion relations corresponding
to the intermediate-energy branch is closed at $q=\rm{\pi}$. Here, we should emphasize that there is no phase transition near this threshold, but still
there is a dramatic change in the nature of the excitations of energy $\propto J_1$.

\subsection{\label{section2-G} Quasi-particles in a truncated Hilbert Space}

Motivated by the results in the preceding sections, we now more formally construct a truncated Hilbert space in which the number of internal
trimer excitations is limited to one doublon or quarton (with no restriction on their internal magnetization).
We could in principle carry out low-order perturbation theory in this space, but the spinons are not easily accounted for in this way.
Here our goal is instead to demonstrate that the spinons, doublons, and quartons can all be present in
the spectrum originating from a severe truncation of the Hilbert space of the trimer states. We  therefore carry out full ED calculations and
construct the dynamic structure factor based on only the approximation of truncated Hilbert space with at most one internal trimer excitation.

\begin{figure}[t]
\includegraphics[width=8cm]{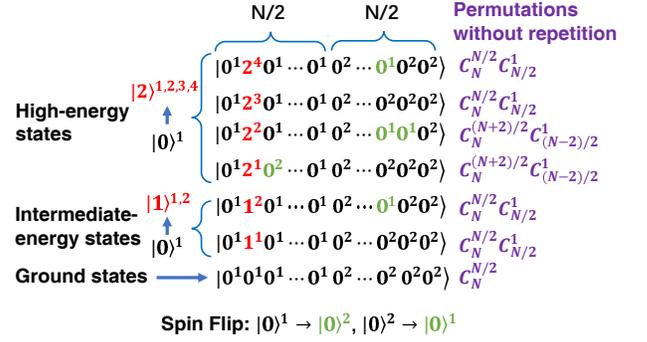}
\caption{\label{figure7} \textbf{Representation of the truncated Hilbert space}. This space contains all the single-trimer eigenstates  (see Fig.~\ref{figure3}) in addition
to the full space of states of the background spin chain, restricted to the total $M=0$ sector.
There are $C_N^{N/2}$ ground  states with $M=0$ formed by permutation without repetition of
$|0^1 \cdots 0^1  0^2 \cdots 0^2 \rangle$. For states containing a doublon or a quarton, one excited trimer (represented by red color)
is present instead of one of the doublet trimer ground states. The excitations
correspond to changes in quantum numbers obeying $|\Delta M|=0$, which in some cases require a flip of a background spin (represented by green color).
Here,  the excitations  $|0\rangle^2 \rightarrow |1\rangle^{1,2} $ and $|0\rangle^2 \rightarrow |2\rangle^{1,2,3,4} $ are not included since a duplicate
of the state can be found from among the already constructed configurations with $| \Delta M|=0$.}
\end{figure}
\begin{figure*}
	\includegraphics[width=16cm]{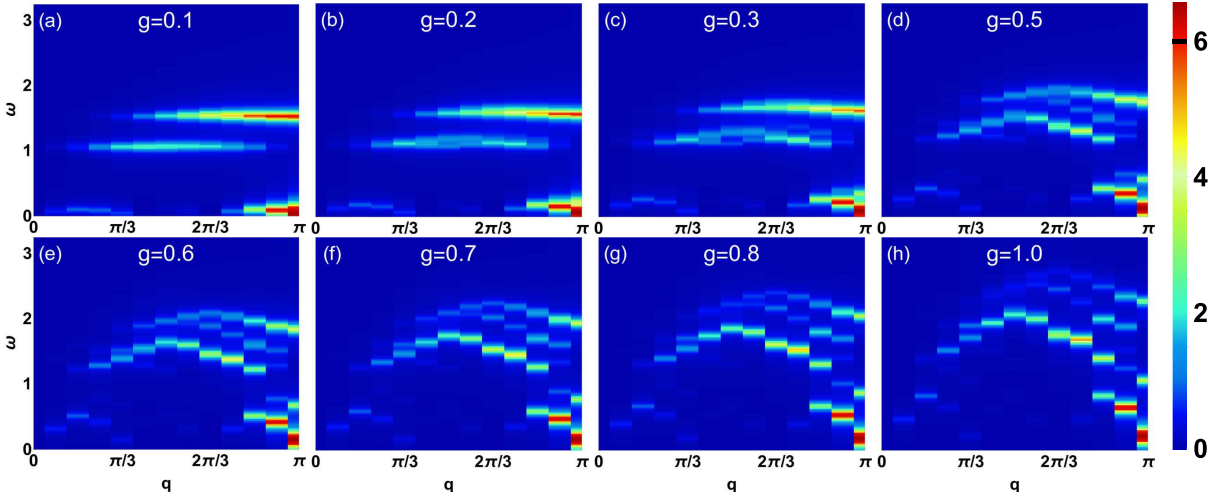}
	\caption{\label{figure8} \textbf{Dynamic spin structure factor $\tilde{\mathcal{S}}^{zz}(q,\omega)$ obtained by ED in the truncated Hilbert space}.  (a)-(h) are the ED calculations for different $g$ values. Here, the system with $L=24$ spins ($N=8$ unit cells) is considered.
		A Lorentzian broadening of width $0.05$ was used of the $\delta$ functions in Eq.~(\ref{newsqwdef}). The boundary between the linear and logarithmic color mapping is
		$U_0=6$ (see definition in the caption of Fig.~\ref{figure2}) as indicated on the color bar.}
\end{figure*}

We construct a  set of states $\Omega$ formed by the eigenvectors of isolated trimers as shown in Fig.~\ref{figure7}. In order to
capture the spinon continuum, we include all $C_N^{N/2}$ combinations of the trimer doublet ground states for which the total magnetic quantum number
$M=0$. For the single excited trimer in its excited state
there are several options for replacing the doublet ground states, as indicated in Fig.~\ref{figure7}. The $\Omega$ covers all the  $M=0$  sates including the spinons, one doublon and one quarton, which is a crucial condition for realizing the full spectrum  in this truncated Hilbert space.
The  states contain
\begin{mathletters}
\begin{equation}
2 C_N^{N/2} C_{N/2}^{1}~~{\rm doublons},
\end{equation}
\begin{equation}
2 \left(C_N^{N/2} C_{N/2}^{1} + C_N^{(N+2)/2} C_{(N-2)/2}^{1}\right)~~{\rm quartons},
\end{equation}
\end{mathletters}
and the total number of states $\phi_i \in \Omega$, including those without internal trimer excitations, is
\begin{eqnarray}
C_N^{N/2}+4 C_N^{N/2} C_{N/2}^{1} + 2 C_N^{(N+2)/2} C_{(N-2)/2}^{1}.
\end{eqnarray}
The effective Hamiltonian matrix in $\Omega$ is
\begin{eqnarray}
\label{transform}
\mathcal{H}_{ij} = \langle \phi_i | H | \phi_j \rangle,
\end{eqnarray}
where the states are visualized in Fig.~\ref{figure7}.

To diagonalize the effective Hamiltonian, we first
Fourier transform the spin operator $S_{q}^z$ [also see Eq.~(\ref{spin transformation}) in Methods~\ref{methods}],
\begin{eqnarray}
	S_{q}^z=\frac{1}{\sqrt{3 N}} \sum_{i=0}^{L-1} \rm{e}^{-i \emph{r}_{i} \emph{q}} \emph{S}_{i}^{z},
\end{eqnarray}
 with a unit cell index $R=0,\cdots,N-1$,
\begin{equation}
  S_{q}^{z}=\frac{1}{\sqrt{3N}}\sum_{R=0}^{N-1} \rm{e}^{-\rm{i} \emph{3Rq}}
   ( \emph{S}_{\emph{R,a}}^{\emph{z}} +  \rm{e}^{-\rm{i} \emph{q}}\emph{S}_{\emph{R,b}}^{\emph{z}} +\rm{e}^{-\rm{i} \emph{2q}} \emph{S}_{\emph{R,c}}^{\emph{z}} ),
\end{equation}
 The spin operators in the truncated Hilbert space are given by
\begin{eqnarray}
\label{new spin}
\left(\mathbb{S}_{R,\alpha}^{z}\right)_{ij} = \langle \phi_i | S_{R,\alpha}^{z} | \phi_j \rangle,
\end{eqnarray}
where the intratrimer labels $\alpha \in \{a,b,c\}$. Then,  above spin operator in momentum space is written as
\begin{equation}
  \mathbb{S}_{q}^{z}=\frac{1}{\sqrt{3N}}\sum_{R=0}^{N-1} \rm{e}^{-\rm{i}\emph{3Rq}}
  ( \mathbb{S}_{\emph{R,a}}^{\emph{z}} +  \rm{e}^{-\rm{i} \emph{q}}\mathbb{S}_{\emph{R,b}}^{\emph{z}} +\rm{e}^{-\rm{i} \emph{2q}} \mathbb{S}_{\emph{R,c}}^{\emph{z}}  ),
\end{equation}
where the momenta is still $q = 2n \rm{\pi}/L, n=0,1,\cdots,L-1$. The dynamic structure factor  is given by
\begin{equation}
\label{newsqwdef}
\tilde{\mathcal{S}}^{zz} (q,\omega)=
  \sum_{m}\left|\left\langle \Psi_m |\mathbb{S}_{q}^{z}|\Psi_0\right\rangle \right|^2 \delta\left[\omega-(\mathbb{E}_m-\mathbb{E}_0)\right],
\end{equation}
where $\left| \Psi_m \right\rangle$ is the $m$th eigenstate with energy $\mathbb{E}_m $.

Results for $L=24$ are already enough to reveal the key features of the excitations and spectral functions.
As shown in Fig.~\ref{figure8}, when $g \leq 0.5$ the spectral shapes and weights coincide well with the results of QMC-SAC (Fig.~\ref{figure2}).
 For $g > 0.5$, the agreement gradually deteriorates, which confirms that the single propagating trimer excitations are no longer a good quasiparticles
of the full system. In particular, the doublon and quarton bands fail to merge into the spinon continuum in the truncated Hilbert space, and this failure also
naturally corresponds to the inability of the truncated Hilbert space to capture the fractionalization of the internal trimer excitations.

It is remarkable
that the agreement with the full and truncated calculations is good up to $g$ as large as $0.5$. The truncated calculation also captures some of the arch feature
of the upper bound of the spinon continuum for $g > 0.5$. This also implies that the fractionalization of the doublons and quartons has not yet set in when
the arch forms. The fractionalization likely sets in only at $g \approx 0.72$, when the arch touches the lower-energy portion of the continuum, and some of
the high-energy excitations may remain confined until $g=1$ (similar to the 2D Heisenberg model, where only some of the high-energy
excitations exhibit signs of fractionalization \cite{SAC2}). At the same time the spectral weight of the low-energy spinons evolves to form the
lower edge of the conventional spinon continuum as $g \to 1$.

\subsection{\label{section2-H} Simplified pictures of doublons and quartons}

\begin{figure}[t]
\includegraphics[width=8cm]{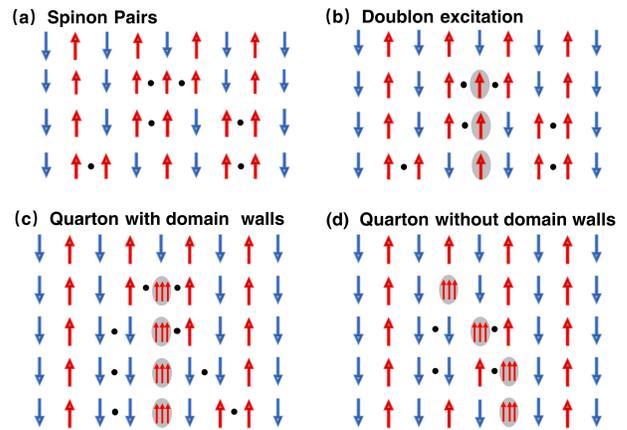}
\caption{\label{figure9} \textbf{Schematic illustrations of propagating doublons and quartons}. In each case (a)-(d), the excitation mechanism and propagation
of a quasiparticle is illustrated from top to bottom. (a) The conventional simplified description of spinons in
the HAC, where the dots indicate domain walls between the two realizations of the antiferromagnetic spin state and these domain walls can move
as a result of flips of a pair of neighboring spins. (b) A doublon excitation, where the trimer in the excited state is indicated by the grey shading.
The spinons (domain walls indicated by dots) should be bound to the excited triplet and can be created and destroyed by vacuum fluctuations. Quartons with (c) and without (d) domain
walls for $|\Delta M| =1$ ($S=1$).}
\end{figure}

Here we develop an intuitive picture of the doublons and quartons, generalizing the standard cartoonish picture
of spinons as domain walls in an antiferromagnet.
In the conventional simplified description of spinons in the HAC,
illustrated in Fig.~\ref{figure9}(a), one starts from a staggered spin configuration (mimicking
the quasi-ordered antiferromagnetic true ground state). Flipping one spin creates
an excitation with $|\Delta M| = 1$, which in the actual spin-rotationally
invariant system corresponds to an excitation carrying spin $S=1$. The left and right misalignments of the flipped spin with respect
to its neighbors can be regarded as two domain walls, and these domain walls can move by two
lattice spacings at a time by flipping pairs of adjacent spins (conserving the
magnetization). These mobile domain walls are the spinons, and once the domain walls have
separated the originally flipped spin has lost its identity and is completely
fractionalized into two independently  propagated spinons.

Let us now extend this cartoon-like picture to the doublon
excitation. As shown in Fig.~\ref{figure9}(b), to create an excitation of this type with $|\Delta M|= 1$,
we again start from an antiferromagnetic spin configuration with three parallel spins,  but now the effective spin in the middle is of the excited type. We can again
think of the misaligned spin configurations as associated with domain walls, and once
these domain walls move away from the central doublet site they look just like standard
spinons. However, they may not completely free, but bound to the still present
central doublet spin. The consequence of the binding is that the central doublet
propagates through the system dressed by spinons, and there should be a large number
of internal modes of this composite excitations, leading
to a band of finite width in energy of these  excitations. In addition to the bound spinons forming the cloud, there should also be
freely propagating spinons coexisting with the propagating central doublet excitations, since
on top of such an excitation a pair of low-energy spinons can be created. However, the
observed dynamic spin structure factor, where the initial excitation is created locally
and later destroyed locally, should be dominated by states with only dressed
central doublet and no free spinons, because of matrix element effects in the same way as
two-spinon processes dominate the structure factor of the standard HAC.

Next we turn to the quarton, which offers several possibilities for
cartoon states with $|\Delta M|=1$ according to the trimer excitations listed in Fig.~\ref{figure3}.
For simplicity we consider those based on the $S^z=3/2$ states, where the effective particles can be
fractionalized or not, as illustrated in Figs.~\ref{figure9}(c) and \ref{figure9}(d). In Fig.~\ref{figure9}(c), we
first replace a spin $S^z_i=-1/2$ by the $S^z=3/2$ triplet state. Then we have $\Delta
M=2$, and to bring this down to $\Delta M=1$ we flip one of the neighbor $S^z=1/2$ spins
down. This creates a domain wall, which can travel away from the excited trimer in
the way discussed above. On the other side of the excited trimer, a domain wall can also
propagate out. Thus, as in the case of the central doublet, Fig.~\ref{figure9}(b), the trimer quarton
will create a cloud of bound spinons surrounding it, with the difference that the
separation between the domain walls is an even number of trimers, instead of an odd number
in the central doublet case.

In Fig.~\ref{figure9}(d), we replace an $S^z_i=1/2$ state by the $S^z=3/2$ triplon
state, which gives us $\Delta M=1$ in a single step. Here we just illustrate how this trimer
excitation can propagate with assistance of virtual spinons. Such processes are also possible with
the central doublet in Fig.~\ref{figure9}(b) and the first quarton in Fig.~\ref{figure9}(c). These processes
represent the motion of the center of mass of the spinon-dressed internal trimer excitations. If
we restore spin-rotation symmetry, the cases depicted in Figs.~\ref{figure9}(c)  and \ref{figure9}(d)
would no longer be separable, and the $|S^z|=1/2$ trimer excitations would also be involved on equal footing.

Clearly the above considerations only provide rough cartoons of the actual
excited states, but in the same way as the simple pictures in Fig.~\ref{figure9}(a) have been
important in forming useful intuition about spinons, these similar pictures for the collective
aspects of the internal trimer excitations are also  useful as an illuminating complement
to the spectral functions and perturbative dispersion relations. In particular, the fact
that these high-energy excitations already contain spinons suggests that they eventually
fractionalize into the conventional HAC spinons by unbinding when $g \to 1$. At the same time,
the internal trimer excitation becomes gradually more ill-defined, involving a larger
number of trimers.

\section{\label{section5} Discussion }

We have investigated the dynamic spin structure factor and the nature of the visible excitations of the spin-$1/2$ antiferromagnetic trimer
chain  by employing the QMC-SAC, ED, and approximate analytical  methods.
We showed that changes in the intertrimer interactions leads to different types of collective excitations
related to the HAC and the internal trimer excitations, and we used a perturbative approach and ED calculation in a  truncated  Hilbert space to confirm
the excitation mechanisms.

When $g$ is small, three well separated excitation branches are present. The low-energy continuum corresponds to the deconfined spinon excitations of an
effective HAC with coupling  $J_{\rm eff}=4 J_2 /9$. In the full BZ of the chain, three such continua are present, with different spectral weight
distributions. When $g=1$, the trimer chain reduces to the conventional HAC with its two-spinon continuum of width $\omega \propto J_1$. We have
investigated the cross-over behavior from one type of spinon to the other kind, as well as the manifestations of the fractionalization of the higher-energy
quasiparticles when they evolve into conventional spinon continuum as $g \to 1$.

When $g  \alt 0.2$, the propagating internal trimer excitations form two weakly dispersive bands, which we named doublons and quartons according to the
nature of the excited isolated trimers. For larger $g$, these two quasiparticle bands merge into each other, and when $g \to 1$ they lose their identity
completely as they fractionalize and evolve into the conventional spinon continuum.
The coexistence of two kinds of emergent spinon branches and two bands of trimer excitations for intermediate $g$ give rise to interesting spectral
signatures. Perturbatively, we identify a threshold value $g_{\rm{t}}=9/(4\rm{\pi}) \approx
0.716$ of the coupling ratio at which these excitations merge into a single continuum, and our numerical results confirm the behavior for $g$ close to
this value. Our calculations in a truncated Hilbert space involving all the spin states of the background chain but only one doublon and one quarton.
Comparing the results with those of the other calculations confirm the stability of the internal quasi-particle excitations up to $g \approx 0.5$,
while for larger $g$ the description of the full system requires a larger number of degrees of freedom to describe the fractionalization process.

It would be very interesting to explore the details of the fractionalization mechanism in future calculations using other theoretical and numerical
approaches. The trimer chain is perhaps the simplest setting in which such mechanisms can be explored---theoretically as well as experimentally. We note that there are already examples of coupled-trimer quantum magnets as discussed in the introduction, for instance,  \ce{A_3Cu_3(PO_4)_4}
(\ce{A=Ca, Sr, Pb}) \cite{PhysRevB.71.144411,drillon19931d,belik2005long,PhysRevB.76.014409}. From the inelastic neutron-scattering spectra measured at $8K$ in  \ce{Pb_3Cu_3(PO_4)_4} \cite{PhysRevB.71.144411},
 two flat excitations at $ \omega \sim 9 meV$ and $ \omega  \sim 13.5 meV$  are observed,  which are also revealed by the intermediate-energy (at $ \omega \sim J_1$) and high-energy (at $ \omega \sim 1.5 J_1$) excitations  in our  theoretical results when $g=0.1$.
  Since the  intertrimer couplings are small in  this material, the trimers are approximately isolated, some features like the energies and relative intensity of two flat bands (see Figs. 2 and 3 in Ref.~\cite{PhysRevB.71.144411}) can also be compared with our results.
    However, the trimers in  \ce{Pb_3Cu_3(PO_4)_4} do not correspond directly to our  linear chain model with only with
nearest-neighbor couplings and $g < 1$.
It is very likely that materials can be synthesized that correspond closely to our model.  Once such a quasi-1D material has
been identified, our results will be helpful for interpreting inelastic neutron scattering and other experiments
probing dynamical properties that beyond the spin waves and conventional spinons,

High-energy ($\sim J$) spin excitations of quantum magnets are less studied than the low-energy modes, but are attracting growing interest
motivated by the  emergence of unusual features in the spectral functions of the high-$T_c$ cuprates \cite{PhysRevLett.105.247001,Zhou2013,Ishii2014,PhysRevB.91.184513,Song2021}, as well as in other antiferromagnets described by the the 2D Heisenberg model \cite{piazza15,SAC2}.
These features have been interpreted as partial fractionalization of magnons, which in the presence of additional interactions can evolve into full
fractionalization in some parts of the BZ. Our results offer useful insights for further exploring coexisting exotic excitations and fractionalization within a relatively simple theoretical framework.

\section{\label{methods}Methods}

\subsection{Stochastic analytic continuation of quantum Monte Carlo data}
In this section, we outline the SAC of QMC data.
The imaginary-time correlation function $G_q(\tau)$ corresponding to $\mathcal{S}(q,\omega)$ is given by
\begin{eqnarray}
	G_q(\tau)=\left\langle S_q (\tau) S_{-q}(0) \right\rangle,
\end{eqnarray}
from which $\mathcal{S}(q,\omega)$ can in principle be reconstructed by inverting the relationship
\begin{eqnarray}
	\label{Gtau}
	G_q(\tau)=\frac{1}{\rm{\pi}} \int_{-\infty}^{\infty} d \omega \mathcal{S} (q,\omega) \rm{e}^{-\tau \omega}.
\end{eqnarray}
In reality, the inversion procedure has limited frequency resolution due to the incomplete information available from QMC calculations of
$G_q(\tau)$ on a finite grid of points and with statistical errors. Nevertheless, with the best available analytical continuation tools and
small statistical errors achievable with long runs using efficient algorithms, quantitatively useful information can be extracted.

In our QMC simulations \cite{SSE1}, we obtain unbiased statistical estimates of $G_q(\tau)$ for a set of imaginary-time points
$\left\{ \tau_i\right\}$. Since the statistical fluctuations for different time points are highly correlated, we also have to compute the covariance
matrix. With a number of QMC data bins $N_B$, based on sufficiently long simulation segments to be in practice uncorrelated, we obtain
the averages $\bar{G}_{q}(\tau_i)=\sum_{b} G_{q}^{b}(\tau_i) / N_{B}$ and the covariance matrix
\begin{eqnarray}
	C_{q}(i,j)=\sum_{b=1}^{N_{B}}\frac{[G_{q}^{b}(\tau_i)-\bar{G}_{q}(\tau_i)]
		[G_{q}^{b}(\tau_j)-\bar{G}_{q}(\tau_j)]}{N_{B}(N_{B}-1)}.~
\end{eqnarray}
In practice, we also normalize the correlation functions so that $\bar{G}_{q}(\tau=0)=1$, which automatically removes the covariance
corresponding to an overall uniform fluctuation of the correlations. The remaining covariance is still significant and has to be taken into account
for the method to be statistically sound.

In the SAC process, $\mathcal{S}(q,\omega)$ is parameterized with a large number of $\delta$-functions. Instead of the commonly
used fixed grid with equally spaced $\delta$-functions with sampled amplitudes, we here use the approach where the $\delta$-functions all have equal
amplitude and instead the frequencies are sampled. The mean density of $\delta$-functions, accumulated in a histogram, then represents the normalized
spectral function $\mathcal{S}(q,\omega)$. The normalization factor  $\bar{G}_{q}(\tau=0)$ is reintroduced after the analytic continuation
so that the correct spectral weight is recovered.

The most complete discussion of the sampling process is currently in Ref.~\cite{SAC2}. Here we just briefly review the variant of the method corresponding to
Fig.~1(a) of Ref.~\cite{SAC2}, which was also used in Refs.~\cite{PhysRevLett.118.147207,PhysRevB.98.174421} where some additional tests and comparisons with
other methods are presented. For a given sampled set of the equal-amplitude $\delta$-functions, the
corresponding imaginary-time function $G^{s}_{q}(\tau_i)$ on the chosen set of points $\{\tau_i\}$ is calculated according to Eq.~(\ref{Gtau}).
The goodness of fit, $\chi^{2}$, defines the closeness to the corresponding QMC data as:
\begin{equation}
	\label{equation6}
	\chi^{2}=\sum_{i,j=1}^{N_{\tau}}\left( G^{s}_{q}(\tau_i)-\bar{G}_{q}(\tau_i)\right)C_{q}^{-1}(i,j)
	\left(G^{s}_{q}(\tau_j)-\bar{G}_{q}(\tau_j)\right ).
\end{equation}
The  spectral function is sampled in a Monte Carlo simulation using the probability distribution
\begin{eqnarray}
	P(\mathcal{S}) \propto \exp \left(-\frac{\chi^{2}}{2 \Theta}\right).
\end{eqnarray}
Here the fictitious temperature $\Theta$ is set so that
\begin{eqnarray}
	\left\langle\chi^{2}\right\rangle \approx \chi_{\min }^{2}+\sqrt{2 \chi_{\min }^{2}},
\end{eqnarray}
which provides a natural scale of fluctuations at which we do not ``fit to the errors'' while a good fit is still automatically guaranteed.

We now turn to the Fourier transform of the spin operator, $S_{q}^z$, for the system with three spins per unit cell. Using the full BZ of the
spin chain of length $L=3N$, we define
\begin{eqnarray}
	\label{spin transformation}
	S_{q}^z=\frac{1}{\sqrt{3 N}} \sum_{i=0}^{L-1} \rm{e}^{-\rm{i} \emph{r}_{i} \emph{q}} \emph{S}_{\emph{i}}^{z},
\end{eqnarray}
where $q=2n \rm{\pi}/L$, $n=0,1,\ldots,L-1$, and the corresponding imaginary-time correlation function is
\begin{eqnarray}
	\label{regular}
	G_{q}(\tau)=3\left\langle S_{q}^{z}(\tau) S_{-q}^{z}(0)\right\rangle.
\end{eqnarray}
It is also useful to consider a reduced BZ. Denoting by $S_{i,\alpha}^{z}$ the spin at position $\alpha \in \{a,b,c\}$ of the $i$th unit cell, we define
\begin{eqnarray}
	S_{q,\alpha}^z=\frac{1}{\sqrt{N}} \sum_{i=0}^{N-1} \rm{e}^{-\rm{i} \emph{r}_{i} \emph{q}} \emph{S}_{\emph{i},\alpha}^{\emph{z}},
\end{eqnarray}
where $q=2n \rm{\pi}/N$, $n=0,1,\ldots,N-1$. The correlation function for the reduced BZ is then assembled as:
\begin{eqnarray}
	\label{reduced}
	G_{q}^{'}(\tau)=\sum_{\alpha}^{a,b,c}\left\langle S_{q,\alpha}^{z}(\tau) S_{-q,\alpha}^{z}(0)\right\rangle.
\end{eqnarray}
In principle we could also consider other form factors internally in the unit cells, but for our purposes here it suffices to consider the reduced BZ with
the uniform summation, along with results for the full BZ based on Eq.~(\ref{regular}).

We have performed the QMC calculation with the length of the spin chain up to $L = 192$ $(N = 64)$. To obtain results representing the low-temperature
limit $T \rightarrow 0$, we scale the inverse temperature $\beta=J_1/T = 4L$. This low temperature is also necessary in order to resolve the spectral
features appearing at very low energies, which are reflected in the imaginary-time correlations at large $\tau$. In the SAC procedure, the statistical noise
of the underlying imaginary-time data is a decisive factor governing the frequency resolution. Normalizing $G_{q}(\tau)$ by setting $G_{q}(0)=1$
as explained above, the statistical errors vanish as $\tau \rightarrow 0$ and approach roughly a constant value as $\tau$ is increased. This almost constant
statistical error for large $\tau$ is a good measure of the level of the statistical errors \cite{SAC1,SAC2}. We have performed sufficiently long calculations
to achieve an error level of approximately $10^{-6}$ for most of the results presented  above.

\section{Data Availability}
The data that support the findings of this study are available from the corresponding authors upon reasonable request.

\section{acknowledgments}
This project is supported by NKRDPC-2017YFA0206203, NKRDPC-2018YFA0306001, NSFC-11974432, NSFC-11804401, NSFC-11832019, GBABRF-2019A1515011337, and Leading Talent Program of Guangdong Special Projects. J.Q.C. is supported by NSFC-12047562. H.Q.W. is also supported by Fundamental Research Funds for the Central Universities, Sun Yat-sen University (Grant No. 2021qntd27). A.W.S. was supported by the NSF under Grant No. DMR-1710170 and by the Simons Foundation under Simons Investigator Grant No.~511064.

\section{Author  Contributions}
J.Q.C. and J.L. contributed equally. D.X.Y., H.Q.W., A.W.S., and J.Q.C. conceived and designed the project. J.L. performed the QMC-SAC simulations.
H.Q.W. and J.Q.C. performed the ED simulations. J.Q.C. analyzed the numerical data under the supervision of H.Q.W., A.W.S. and D.X.Y. All authors
contributed to the interpretation of the results and wrote the paper.

\section{Competing Interests}
The authors declare no competing interests.

{\flushleft \bf REFERENCES}

{\flushleft \bf ADDITIONAL INFORMATION}
\vspace{-1.0mm}
{\flushleft \bf Supplementary Information} is available for the paper at this weblink

{\flushleft \bf Correspondence} and requests for materials should be addressed to H.Q.W., A.W.S. or D.X.Y.

\clearpage
\begin{center}
{\centering \bf Supplementary Note 1: Exact solutions of small systems}
\end{center}
\vspace{1.5mm}

In addition to the QMC-SAC calculations, we have also applied the ED method to obtain reliable results. ED is the most basic numerical method for determining the eigenvalues
and eigenstates of a quantum magnet with up to tens of spin-$1/2$ spins. If all the eigenstates have been obtained for a very small system,
Eq.~(3) in main text can be computed directly, and somewhat larger systems can be considered within the Lanczos ground-state method combined with
the continued fraction expansion \cite{dagotto1996}. To reach larger sizes,  we can further use symmetries to block diagonalize the Hamiltonian and reduce
the computation time and memory requirement. Although  ED is  limited to extracting exact information from relatively small systems, where finite-size effects
are still significant, we can   gain more reliable insights pertaining to the thermodynamic limit by  comparing results from the two different approaches.

In Supplementary Figure ~\ref{figure10}, we present the spin excitation spectra of the trimer chain obtained by ED calculations for different values of the coupling ratio $g$. Here the ED results are calculated from a chain with $30$ spins. We have  applied broadening in the momentum direction to aid in the visualization of spectrum.

\begin{figure*}
	\includegraphics[width=16cm]{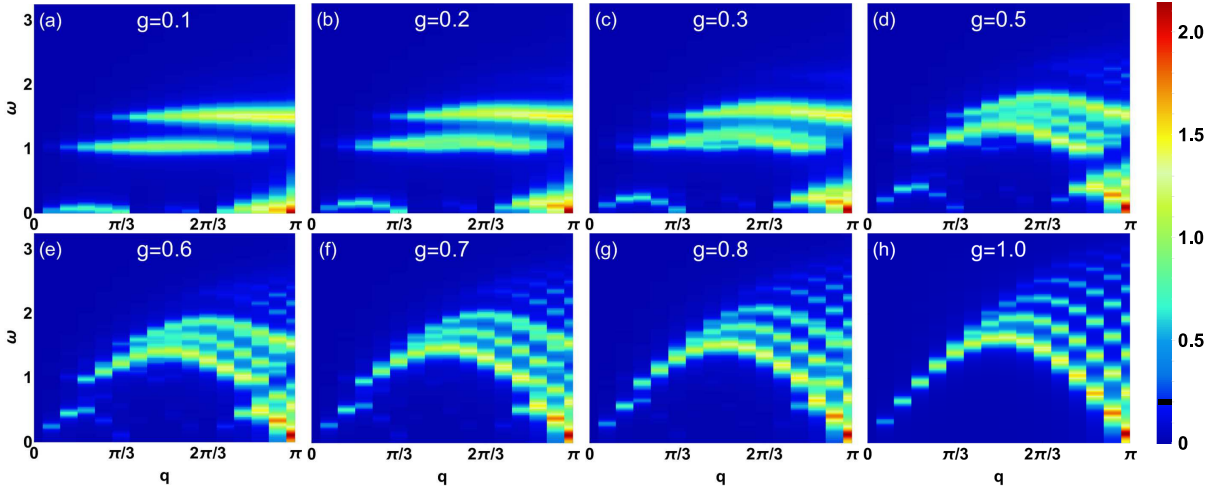}
	\caption{\label{figure10} Dynamic spin structure factor $\mathcal{S}(q,\omega)$ obtained by ED for the trimer chain with $L=30$ spins ($N=10$ unit cells)
		at different coupling ratios $g$. The boundary between the linear and logarithmic color mappings is $U_0=0.2$, as indicated on the color bar.  In this ED calculation a Lorentzian broadening factor $0.05$ was used of the $\delta$ functions in Eq.~(3) of main text, and an uniform stretching of the 15 points in the
		momentum direction of the half BZ shown here was also applied by the software used to construct the graph.}
\end{figure*}
\vspace{1.5mm}

In the ED calculations we can clearly see fine structure due to the small number of $\delta$-functions (which are broadened for visualization) in Eq.~(3) of main text.
The visible bands formed by the individual excitations of the small chain fall within the spinon continuum forming in the thermodynamic limit. The ED and QMC-SAC spectral weight distributions are similar, but the QMC-SAC exhibit
the full spinon continua more clearly. The two distinct excitation branches above the spinon continuum are very similar in the two sets of results.

\vspace{1.5mm}

\begin{center}
{\bf \label{effective coupling} Supplementary Note 2: Effective Heisenberg coupling}
\end{center}
\vspace{1.5mm}

To derive the effective HAC arising from the doublet trimer ground states, we formally split the original Hamiltonian into three-spin blocks and apply
the generic Kadanoff method \cite{drell1977quantum,jullien1978zero,RealSpace1996,PhysRevA.77.032346}. The trimer chain Hamiltonian is thus decomposed
into the intratrimer ($H^{\rm{t}}$) and intertrimer ($H^{\rm{tt}}$) parts. The lowest eigenstates of each trimer are then used to construct
the basis for the effective Hilbert space.  To achieve an effective Hamiltonian with structure similarity to the original one, the full Hamiltonian is
projected onto the effective Hilbert space. The intratrimer Hamiltonian, which only contains the $J_1$ couplings, reads
\begin{equation}
	H^{\rm{t}} =  J_1 \sum_{j=1}^{N} \left(\mathbf{S}_{j,a} \cdot \mathbf{S}_{j,b} +\mathbf{S}_{j,b} \cdot \mathbf{S}_{j,c} \right),
\end{equation}
where $N $ represents the number of trimers and we again refer to Fig.~1 of main text for the definition of the intratrimer labels $a$, $b$, $c$.
The intertrimer Hamiltonian is given by
\begin{equation}
	H^{\rm{tt}}=J_2 \sum_{j=1}^{N-1}\mathbf{S}_{j,c} \cdot \mathbf{S}_{j+1,a},
\end{equation}
which only contains the interaction between the $c$ site of $j$th trimer and the $a$ site of $(j+1)$th trimer. Each trimer has two degenerate ground states (see Fig.~3 in main text);
\begin{eqnarray}
	\left|0\right\rangle^{1}  = \frac{1}{{\sqrt 6 }}\left( {\left| { \uparrow  \downarrow  \downarrow } \right\rangle  - 2\left| { \downarrow  \uparrow  \downarrow } \right\rangle  + \left| { \downarrow  \downarrow  \uparrow } \right\rangle } \right),\\
	\left|0\right\rangle^{2}  = \frac{1}{{\sqrt 6 }}\left( {\left| { \downarrow  \uparrow  \uparrow } \right\rangle   - 2\left| { \uparrow  \downarrow  \uparrow } \right\rangle  +  \left| { \uparrow  \uparrow  \downarrow } \right\rangle} \right),	
\end{eqnarray}
where $ \left|\uparrow \right\rangle $ and $\left|\downarrow \right\rangle$ are the eigenstates of the spin operator $\mathbf{\sigma}_z$. The projection operator of the
$j$th trimer,
\begin{equation}
	\mathbf{P}_j= \left|0 \right\rangle^{2}_j  \left \langle \Uparrow  \right|_j + \left|0 \right\rangle^{1}_j \left\langle \Downarrow \right|_j
\end{equation}
is built to project the Hamiltonian onto the low-energy subspace, where $\left|  \Uparrow  \right\rangle$ and $\left|  \Downarrow  \right\rangle$ are
the renamed base kets in this effective Hilbert space. The effective Hamiltonian up to the first-order correction is
\begin{eqnarray}
	\label{Heff}
	H_{\rm{eff}} = {\mathbf{P}^\dag }H^{\rm{t}}\mathbf{P} +  {\mathbf{P}^\dag }  H^{\rm{tt}}\mathbf{P},
\end{eqnarray}
where the total projection operator is $\mathbf{P} = \prod_{j=1}^{N} \mathbf{P}_j $. The effective Pauli operators are given by
\begin{equation}
	\widetilde{\mathbf{\sigma}}_j^{\beta}= \xi_i^{\beta} \mathbf{P}_j^\dag \mathbf{\sigma}_{i,j}^{\beta} \mathbf{P}_j,~~ ( i=a,b,c;~ \beta=x,y,z),
\end{equation}
where  $\xi_i^{\beta}$ is the coefficient required to preserve the SU(2) algebra.  Inserting the effective Pauli operators into Eq.~(\ref{Heff}), we
obtain the effective Hamiltonian
\begin{equation}
	H_{\rm{eff}}=J_{\rm eff} \sum_{j=1}^{N} \widetilde{\mathbf{S}}_{j} \cdot \widetilde{\mathbf{S}}_{j+1},
\end{equation}
which describes an isotropic HAC in which the effective coupling strength only depends on the intertrimer coupling, $J_{\rm eff} = 4J_2/9$.

\begin{center}
{ \bf Supplementary Note 3: Dispersion relations \label{DR}}
\end{center}
\vspace{1.5mm}

As we have discussed in the main text, the internal excitations of the trimer chain are almost localized when $g$ is small, and cannot be classified as  magnons or triplons.
In order to obtain further insight into the nature of these excitations, we propose a scheme to calculate their dispersion relations. The results are shown in
Fig.~4 of main text as a number of dispersion relations corresponding to propagation of internal trimer excitations in the background of a chain with defects
mimicking the complex ground state of the effective HAC. The overall good agreement with the QMC--SAC results on the location of these excitations and their
band widths suggest that our picture of the excitation is correct even though the calculation involves a very rough approximation of the ground state
and a simple ansatz for its excitations.

We begin by assuming that the ground-state wave function of spin chain is a product state of ground states of each trimer,
\begin{equation}
	\label{ground state}
	\left| {\psi}_{\rm{g}}\right\rangle=\left| {0}\right\rangle_1 \left| {0}\right\rangle_2  \cdots \left| {0}\right\rangle_r \cdots \left| {0}\right\rangle_{N},
\end{equation}
where $\left|0\right\rangle$ is one of the two ground states of a single trimer. Thus, we neglect the interactions that cause the effective HAC with its
spinon continuum, and instead deal with $2^N$ degenerate ground states without enforcing any quantum numbers (spin or momentum). Note that we do not apply
degenerate perturbation theory here, because we are targeting excitations above the $2^N$-fold degenerate ground-state manifold. We also do not carry out a
formal perturbation expansion, but construct intuitive variational states including the internal excitations of one trimer.

If the $r$th trimer is excited
from its ground state $\left|0\right\rangle$ to one of its higher states $\left|1\right\rangle$, then the wave function with this localized excitation reads
\begin{equation}
	\left| {\psi_{\rm{e}}}\right\rangle_r=\left| {0}\right\rangle_1 \left| {0}\right\rangle_2  \cdots \left| {1}\right\rangle_r \cdots \left| {0}\right\rangle_N.
\end{equation}
We can give this excitation a momentum and calculate its dispersion relation. Since the trimer has two types of excitations (see Fig.~3 of main text), we will
obtain two bands. Next, we first consider the lower $\omega=J_1$ doublon excitation of the isolated trimer, and then defer the similar but
more complicated case of the  $\omega=3J_1/2$ quarton excitation.

\begin{figure*}[t]
	\includegraphics[width=12cm]{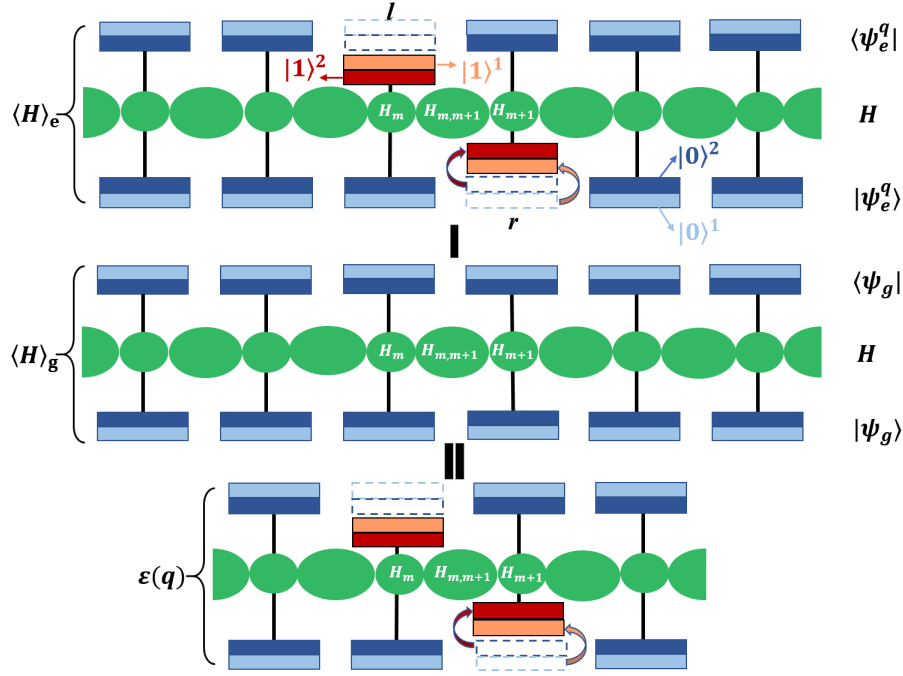}
	\caption{\label{suppfigure} Graphical representation of the calculation of the dispersion relation
		$\epsilon(q)=\left\langle H \right\rangle_{\rm{e}} - \left\langle H \right\rangle_{\rm{g}}$ of the intermediate-energy
		excitation within our approximation. The doubly-degenerate trimer eigenstates  are shown as lighter and darker blue (ground states)
		and orange or red (first excited states). With these states, the intermediate-energy excitations with $|\Delta M|=1$ originate from a trimer ground state
		$\left|0\right\rangle^{1}_r$ on the trimer located at $r$ when excited to $\left|1\right\rangle^{2}_r$ or $\left|0\right\rangle^{2}_r$ excited to
		$\left|1\right\rangle^{1}_r$ (and in the corresponding bra states we use the site index $l$ instead of $r$). The surrounding trimers, which can be in
		either of their doubly-degenerate ground states, are not affected. The excitations are given momentum $q$, and the matrix elements contributing
		to the dispersion relation are indicated. In addition to the case $r=l+1$ shown, $r=l-1$ contributes equally and $r=l$ gives a different contribution.
		No other relationships between $r$ and $l$ contribute.}
\end{figure*}

\subsection{\label{DRint}Intermediate-energy doublon excitation}

For simplicity, we first consider the approximate uniform ground-state wave function with all the trimers in the same ground state of the
isolated trimer,
$\left|0\right\rangle^{2}=(\left|\downarrow \uparrow \uparrow \right\rangle -2\left|\uparrow \downarrow \uparrow \right\rangle+\left|\uparrow \uparrow \downarrow \right\rangle)/\sqrt{6}$.
This state clearly has momentum $q=0$. In the lowest excitation involving the internal trimer excitation, one of the trimers is put in its first excited
state $\left|1\right\rangle^{1}=(\left| \downarrow \downarrow \uparrow \right\rangle-\left|\uparrow \downarrow \downarrow \right\rangle)/\sqrt{2}$
with $\Delta M =-1$. Fourier transforming such a state, the one-trimer excited state in momentum space is given by
\begin{equation}
	\left| {\psi_{\rm{e}}^q }\right\rangle=\frac{1}{\sqrt{N}} \sum_{r=1}^{N} \rm{e}^{-\rm{i} q \emph{r}} \left| {\psi_{\rm{e}}}\right\rangle_r.
\end{equation}

The total Hamiltonian with periodic boundary conditions is the summation of intratrimer Hamiltonian $H_m$ and the intertrimer
Hamiltonian $H_{m,m+1}$
\begin{equation}
	H=\sum_{m=1}^{N} H_{m}+\sum_{m=1}^{N} H_{m, m+1}.
\end{equation}
Next, we calculate the expectation values of this Hamiltonian in the ground state and the first excited trimer momentum state. For the ground state of trimer chain, the expectation value of $H$ trivially evaluates to
\begin{equation}
	\left\langle H \right\rangle_{\rm{g}}=\left\langle \psi_{\rm{g}}\right|H \left| \psi_{\rm{g}}\right\rangle
	= N E_0 +\frac{N}{9} J_2.
\end{equation}
For the one-trimer excited state, we consider separately the two terms of the expectation value of $H$ in momentum space:
$\left\langle \psi^q_{\rm{e}}\right|\sum_{m=1}^{N}H_m \left| \psi^q_{\rm{e}} \right\rangle$ and $ \left\langle \psi^q_{\rm{e}} \right| \sum_{m=1}^{N}H_{m,m+1}\left| \psi^q_{\rm{e}} \right\rangle $. The first term only contributes the gap  $E_1-E_0$. To evaluate the second term,
we let $H_{m,m+1}$ act on the $c$ spin of  $m$th trimer and the $a$ spin of the $(m+1)$th trimer. The excited trimers labeled by $l$ and $r$ in the bra
$\left\langle \psi^{q}_{\rm{e}}\right|$ and ket $\left| \psi^{q}_{\rm{e}}\right\rangle$ should take all the values of $\left[1,N\right]$.
Practically, only the cases $l=r$ and  $l=r\pm 1$ contribute, however, because $ \left\langle \psi^{q}_{\rm{e}}\right|$ and
$\left| \psi^{q}_{\rm{e}}\right\rangle$ are orthogonal for $|l-r|\geq 2$. Therefore, the expectation value of $H$ for the one-trimer excited
state is
\begin{eqnarray}
	\left\langle H \right\rangle_{\rm{e}}&=&\left\langle \psi^{q}_{\rm{e}}\right|H \left| \psi^{q}_{\rm{e}}\right\rangle \nonumber \\
	=&& \frac{1}{N} \sum_{r=1}^{N} \sum_{l=1}^{N} \rm{e}^{-\rm{i} q \emph{r}+\rm{i} q \emph{l}}
	\left\langle 0^{2} \cdots 1_{\emph{l}}^{1}  \cdots 0^{2} \right| \sum_{m=1}^{N}H_m \nonumber \\
&&+\sum_{m=1}^{N}H_{m,m+1} \left| 0^{2} \cdots 1_{\emph{r}}^{1} \cdots 0^{2} \right\rangle  \nonumber\\
	&=& (N-1) E_0 + E_{1}+J_2 \left( \frac{N-2}{9}- \frac{1}{3} \cos{q}\right).~~
\end{eqnarray}
Thus, the dispersion relation is of what we refer to as the intermediate-energy excitation in reduced BZ is
\begin{eqnarray}
	\label{intermediate}
	\epsilon(q)&=&\left\langle H \right\rangle_{\rm{e}} - \left\langle H \right\rangle_{\rm{g}} \nonumber\\
	&=& E_1 - E_0  - \frac{2}{9}J_2 -\frac{1}{3}J_2\cos{q},
\end{eqnarray}
independently of the length of the spin chain.
Unfolding this dispersion relation by replacing $q$  with $3q$, we can obtain the dispersion relation in full BZ
\begin{eqnarray}
	\epsilon(q)= E_1 - E_0  - \frac{2}{9}J_2 -\frac{1}{3}J_2\cos (3{q}).
\end{eqnarray}
The first two terms describe the gap of the localized excited trimer, and the last two terms reveal the dynamical
features dominated by the intertrimer interaction. By symmetry, for the  excitation from ground state $\left|0\right\rangle^{1}$ to the other first-excited
state $\left|1\right\rangle^{2}$ with $\Delta M=1$, the same dispersion relation, Eq.~(\ref{intermediate}), is also obtained in the reduced BZ.

To better mimic the true many-body ground state, we can consider random arrangements of the states $|0\rangle^1_i$ and $|0\rangle^2_i$ on each site. Exciting
from all possible state in this degenerate manifold will lead to a band of excitations that approximate the true bands arising from combinations of an internal
trimer excitation and spinons. We always start from momentum $0$ in the ground state, and in the excited state we consider the same configuration of
$|0\rangle^1_i$ and $|0\rangle^2_i$ but with one of these trimer ground states at site $r$ excited to the first trimer excitation
$|1\rangle_r^1$ or $|1\rangle_r^2$
such that $|\Delta M|=1$, and this state is given momentum $q$. Because of the form of the Hamiltonian, and examining the energy calculation as
illustrated in Supplementary Figure ~\ref{suppfigure}, we can conclude that the dispersion relations are mainly dependent on the excited trimers and their neighbours. Translated
to finite size, this means that the correct generic result is already obtained from a system with $N=4$ trimers. Thus, we can generate 16 different
dispersion relations for the intermediate-energy excitation, and because of symmetries there are only four different ones. In addition to the one
already given in Eq.~(\ref{intermediate}), the other three are the Eqs.~(6)-(8) in main text.
Although all possibilities have been considered above, the true ground state for  $g>0$ is a total-spin singlet satisfying $M=\sum_{i} M_i =0$, and the numbers
of $\left|0\right\rangle^{1}$ and $\left|0\right\rangle^{2}$ used when forming ground states should therefore be equal. After considering this restriction,
the result of Eq.~(\ref{intermediate}) is eliminated  since its approximation corresponds to a maximal-$M$ uniform  ground state. Only $6$ dispersion
relations described by  Eqs.~(6)-(8) in main text are left for $N=4$ trimers. Each equation corresponds to two degenerate dispersion
relations. All three dispersion relations are graphed in Fig.~5(a) of main text for $g=0.1$. Two of the above dispersion relations are independent of
$q$ since the excitations are localized. Only the $l = r$ terms in Supplementary Figure ~\ref{suppfigure} contribute to these excitations, while the $l=r\pm 1$ terms
lead to propagation of the internal trimer excitations. According to the  characters of this excitation starting from the $g=0$ limit, we refer to it
as the doublon.
\subsection{\label{DRhigh}Dispersion relations of the  high-energy excitation}

Calculating the dispersion of the high-energy excitations evolving from the trimer quartet at $\omega = 3J_1/2$ is more complicated but we follow the
same strategy as for the doublon. In the latter case, the intermediate-energy band is formed by reconfiguring the singlet bond in the original ground state of
the trimer in Fig.~3 of main text at a cost of $J_1$. The higher-energy quarton band is formed out of trimers excited into their $S=3/2$ highest state, which
can be seen as the singlet of two spins excited into a triplet, at a cost of $3J_1/2$. This rough estimation of the excited energy matches very well the
numerical ED and QMC-SAC results for small $g$ values.

Again our calculation does not conserve the total spin of the collective many-body state, but we consider the excitations corresponding to $\Delta S=1$ on
one trimer. In this calculation, we not only need to prepare the ground-state wave functions, but also select the excited states from the quadruplet.
Here, $N=4$ trimers are also adequate to do the calculation at our level of linear-in-$g$ approximation. With each of the $l$th and $r$th excited trimers
having four possible states, at least $2^4$ ground states and $ 4^2$ higher-lying excited states are needed, and there are $2^4\times 4^2=256$ cases in total.
After implementing the restriction of the ground-state wave function having $M =0$, $96$ cases are left. As a result, the dispersion relations  of the  high-energy excitation  in the reduced BZ  fall into $33$ different cases and are displayed in Eqs.~(9)-(17) in the main text.
\\

\begin{center}
{\flushleft \bf Supplementary References}
\end{center}
\vspace{+1.5mm}

\end{document}